# Construct SLOCC invariants via square matrix

Xin-Wei Zha

School of Science, Xi'an Institute of Posts and Telecommunications, Xi'an, 710121, P R China

Abstract  We define a square matrices, by which some stochastic local operations and classical communication (SLOCC) invariants can be obtained. The relation of SLOCC invariants and character polynomial of square matrix are given for three and four qubit states.

## 1. Introduction

Quantification of quantum entanglement is one of the most important problems in quantum information theory. Therefore, it is important to find ways of classifying and quantifying the entanglement properties of quantum states. Invariants under stochastic local operations and classical communication(SLOCC), have been extensively studied in this context[1–13].Recently, Li *et al.* [14]proposed a method of classifying *n*-qubit states into stochastic local operations and classical communication inequivalent families in terms of the rank of the square matrix. According to this inspiration, a improved square matrix definition is put forward, by which some SLOCC invariants can be obtained. Furthermore, the relation of SLOCC invariants and character polynomial of square matrix is given for three and four-qubit states.

## 2. SLOCC invariant and square matrix

In a recent paper [Phys. Rev. A **91**, 012302, (2015)], Li *et al.* proposed a definition of square matrices, which can be defined as[14]

$$Q^{(n)}_{q_1,q_2,\cdots,q_i} = C_{q_1,q_2,\cdots,q_i} (i\sigma_y)^{\otimes(n-i)} \left[ C_{q_1,q_2,\cdots,q_i} \right]^T \tag{1}$$

Where $C_{q_1,q_2,\cdots,q_i}$ is the coefficient matrix and $\sigma_y$ is the Pauli operator.

In the follows, we will give the revise definition of the square matrices. It is known that two states $|\psi\rangle$ and $|\varphi\rangle$ are equivalent under SLOCC if and only if there exist invertible local operators $M_A, M_B, M_C \cdots$, and such that

$$|\psi\rangle_{ABC\cdots} = A_A \otimes A_B \otimes A_C \cdots \otimes |\varphi\rangle_{ABC\cdots} \tag{2}$$

where $A_k \in SL(2)$ are SLOCC operations with $\det(A_k) = 1$.

In an analogous way, we define a square matrices, which can be expressed as

$$F^n_{1,2,\cdots i}(\psi) = (v)^{\otimes(i)} C^n_{1,2,\cdots i}(\psi)(v)^{\otimes(n-i)} \left[ C^n_{1,2,\cdots i}(\psi) \right]^T \tag{3}$$

where $v = i\sigma_y$ and $\sigma_y$ is the Pauli operator.

Invoking the fact that $A_k^T v A_k = \det A_k v$, for $\det A_k = 1$, $A_k^T v A_k = v$ we may have

$$F^n_{1,2,\cdots i}(\psi') = (v)^{\otimes(i)} C^n_{1,2,\cdots i}(\psi')(v)^{\otimes(n-i)} \left[ C^n_{1,2,\cdots i}(\psi') \right]^T \tag{4}$$

using $C_{1,2,\cdots i}^n(\psi') = (A_1 \otimes \cdots A_i) C_{1,2,\cdots i}^n(\psi)(A_{i+1} \otimes \cdots A_n)^T$ (5)

We can obtain $F_{1,2,\cdots i}^n(\psi') = [(A_1 \otimes \cdots A_i)^T]^{-1} F_{1,2,\cdots i}^n(\psi)(A_1 \otimes \cdots A_i)^T$

(6)

Therefore, the square matrices $F(\psi)$ and $F(\psi')$ is similar transformation'

Let the character polynomial of $F(\psi)$ is given by

$$\det\{F(\psi) - \lambda I\} = a_m \lambda^m + \cdots + a_1 \lambda + a_0 \quad (7)$$

This means that $f(\lambda) = \det\{F(\psi) - \lambda I\}$ is a SLOCC invariant polynomial and every $\lambda$ coefficient of $a_i$ is a SLOCC invariant.

3. **The relation of SLOCC invariant with square matrix for three-qubit states**

An arbitrary pure state $|\psi\rangle$ of the three qubits is expressed in the form of

$$|\psi\rangle_{123} = a_0|000\rangle + a_1|001\rangle + a_2|010\rangle + a_3|011\rangle + a_4|100\rangle + a_5|101\rangle + a_6|110\rangle + a_7|111\rangle$$

(8)

It is well known that **the** SLOCC invariant is[4,15]

$$F_1^{(3)} = \langle \psi^*|\hat{\sigma}_{Ay}\hat{\sigma}_{By}\hat{\sigma}_{Cx}|\psi\rangle^2 + \langle \psi^*|\hat{\sigma}_{Ay}\hat{\sigma}_{By}\hat{\sigma}_{Cz}|\psi\rangle^2 - \langle \psi^*|\hat{\sigma}_{Ay}\hat{\sigma}_{By}|\psi\rangle^2$$
$$= 4[(a_0 a_7 - a_1 a_6 - a_2 a_5 + a_3 a_4)^2 + 4(a_0 a_3 - a_1 a_2)(a_5 a_6 - a_4 a_7)] \quad (9)$$

A convenient, and physically significant, choice is the 3-tangle identified by Coffman, Kundu and Wootters[16] : $\tau_{123}^2$, $\tau_{123} = |F_1^{(3)}|$ (10)

For three qubit pure state, the coefficient matrix can be defined

$$C_{1(23)} = \begin{pmatrix} a_0 & a_1 & a_2 & a_3 \\ a_4 & a_5 & a_6 & a_7 \end{pmatrix}, C_{2(13)} = \begin{pmatrix} a_0 & a_1 & a_4 & a_5 \\ a_2 & a_3 & a_6 & a_7 \end{pmatrix}, C_{3(21)} = \begin{pmatrix} a_0 & a_2 & a_4 & a_6 \\ a_1 & a_3 & a_5 & a_7 \end{pmatrix} \quad (11)$$

From [3,11], we have

$$F_{1(23)}(\psi) = \begin{pmatrix} 0 & 1 \\ -1 & 0 \end{pmatrix} \begin{pmatrix} a_0 & a_1 & a_2 & a_3 \\ a_4 & a_5 & a_6 & a_7 \end{pmatrix} \begin{pmatrix} 0 & 0 & 0 & 1 \\ 0 & 0 & -1 & 0 \\ 0 & -1 & 0 & 0 \\ 1 & 0 & 0 & 0 \end{pmatrix} \begin{pmatrix} a_0 & a_4 \\ a_1 & a_5 \\ a_2 & a_6 \\ a_3 & a_7 \end{pmatrix}$$

$$= \begin{pmatrix} a_0 a_7 - a_1 a_6 - a_2 a_5 + a_3 a_4 & 2(a_4 a_7 - a_5 a_6) \\ -2(a_0 a_3 - a_1 a_2) & -(a_0 a_7 - a_1 a_6 - a_2 a_5 + a_3 a_4) \end{pmatrix} \quad (12)$$

The character polynomial of $F_{1(23)}(\psi)$ is given by

$$\det\{F_{1(23)}(\psi) - \lambda I\} = a_2\lambda^2 + a_1\lambda + a_0 \quad (13)$$

Using [12,13], we have the SLOCC invariant

$$a_2 = 1$$

$$a_1 = 0$$

$$a_0 = -(a_0a_7 - a_1a_6 - a_2a_5 + a_3a_4)^2 - 4(a_0a_3 - a_1a_2)(a_5a_6 - a_4a_7)$$
$$= -\frac{1}{4}F_1^{(3)} \quad (14)$$

Therefore, the determinant of the square matrix $F_{1(23)}(\psi)$ is **the SLOCC invariant**.

4. **The relation of SLOCC invariant with square matrix for four-qubit states.**

The states of a four-qubit system can be generally expressed as

$$|\psi\rangle_{1234} = a_0|0000\rangle + a_1|0001\rangle + a_2|0010\rangle + a_3|0011\rangle$$
$$+ a_4|0100\rangle + a_5|0101\rangle + a_6|0110\rangle + a_7|0111\rangle$$
$$+ a_8|1000\rangle + a_9|1001\rangle + a_{10}|1010\rangle + a_{11}|1011\rangle$$
$$+ a_{12}|1100\rangle + a_{13}|1101\rangle + a_{14}|1110\rangle + a_{15}|1111\rangle \quad (15)$$

We know there are four independent algebraic SLOCC invariants for a 4-qubit system[6,8], that is $(H, L, M, D_{xt})$. Where

$$H = 2(a_0a_{15} - a_1a_{14} - a_2a_{13} + a_3a_{12} - a_4a_{11} + a_5a_{10} + a_6a_9 - a_7a_8) \quad (16)$$

and H is a degree-2 invariant whose each term involves only two coefficients.
L and M are degree-4 invariant, are given by the determinants of matrices:

$$L = \begin{vmatrix} a_0 & a_4 & a_8 & a_{12} \\ a_1 & a_5 & a_9 & a_{13} \\ a_2 & a_6 & a_{10} & a_{14} \\ a_3 & a_7 & a_{11} & a_{15} \end{vmatrix}, \quad (17)$$

$$M = \begin{vmatrix} a_0 & a_8 & a_2 & a_{10} \\ a_1 & a_9 & a_3 & a_{11} \\ a_4 & a_{12} & a_6 & a_{14} \\ a_5 & a_{13} & a_7 & a_{15} \end{vmatrix}, \quad (18)$$

$D_{xt}$ is a degree-6 invariants[8], and can be expressed as the determinants of three $3\times 3$ matrices:

$$D_{xt} = \begin{vmatrix} a_0a_6-a_2a_4, & a_0a_7+a_1a_6-a_2a_5-a_3a_4, & a_1a_7-a_3a_5, \\ a_0a_{14}+a_6a_8, & a_0a_{15}+a_6a_9+a_7a_{14}+a_7a_8, & a_1a_{15}+a_7a_9, \\ -a_2a_{12}-a_4a_{10}, & -a_2a_{13}-a_3a_{12}-a_4a_{11}-a_5a_{10}, & -a_3a_{13}-a_5a_{11}, \\ a_8a_{14}-a_{10}a_{12}, & a_8a_{15}+a_9a_{14}-a_{10}a_{13}-a_{11}a_{12}, & a_9a_{15}-a_{11}a_{13}, \end{vmatrix} \quad (19)$$

The coefficient matrix of four-qubit system can be expressed as

$$C_{1(234)} = \begin{pmatrix} a_0 & a_1 & a_2 & a_3 & a_4 & a_5 & a_6 & a_7 \\ a_8 & a_9 & a_{10} & a_{11} & a_{12} & a_{13} & a_{14} & a_{15} \end{pmatrix},$$

$$C_{2(134)} = \begin{pmatrix} a_0 & a_1 & a_2 & a_3 & a_8 & a_9 & a_{10} & a_{11} \\ a_4 & a_5 & a_6 & a_7 & a_{12} & a_{13} & a_{14} & a_{15} \end{pmatrix},$$

$$C_{3(124)} = \begin{pmatrix} a_0 & a_1 & a_4 & a_5 & a_8 & a_9 & a_{12} & a_{13} \\ a_2 & a_3 & a_6 & a_7 & a_{10} & a_{11} & a_{14} & a_{15} \end{pmatrix},$$

$$C_{12(34)} = \begin{pmatrix} a_0 & a_1 & a_2 & a_3 \\ a_4 & a_5 & a_6 & a_7 \\ a_8 & a_9 & a_{10} & a_{11} \\ a_{12} & a_{13} & a_{14} & a_{15} \end{pmatrix}, \quad C_{13(24)} = \begin{pmatrix} a_0 & a_1 & a_4 & a_5 \\ a_2 & a_3 & a_6 & a_7 \\ a_8 & a_9 & a_{12} & a_{13} \\ a_{10} & a_{11} & a_{14} & a_{15} \end{pmatrix},$$

$$C_{14(23)} = \begin{pmatrix} a_0 & a_2 & a_4 & a_6 \\ a_1 & a_3 & a_5 & a_7 \\ a_8 & a_{10} & a_{12} & a_{14} \\ a_9 & a_{11} & a_{13} & a_{15} \end{pmatrix}$$

(20)

From [3,20], we have

$$F_{1(234)}(\psi) = v_1 C_{1(234)}(v_2 \otimes v_3 \otimes v_4)\left[C_{1(234)}\right]^T = \begin{pmatrix} -\frac{1}{2}H & 0 \\ 0 & -\frac{1}{2}H \end{pmatrix}$$

(21)

Therefore, The character polynomial of $F_{1(234)}(\psi)$ is given by

$$\lambda^2 + H\lambda + \frac{1}{4}H^2 \tag{22}$$

Obviously, this character polynomial determine the degree-2 invariant H.

Similarly, we have

$$F_{12(34)}(\psi) = (v_1 \otimes v_2)C_{12(34)}(v_3 \otimes v_4)\left[C_{12(34)}\right]^T$$

$$= \begin{pmatrix} a_0 a_{15} - a_1 a_{14} - a_4 a_{11} + a_5 a_{10} & a_4 a_{15} - a_5 a_{14} - a_6 a_{13} + a_7 a_{12} & a_8 a_{15} - a_9 a_{14} - a_{10} a_{13} + a_{11} a_{12} & 2(a_{12}a_{15} - a_{13}a_{14}) \\ -a_0 a_{11} + a_1 a_{10} + a_2 a_9 - a_3 a_8 & -a_4 a_{11} + a_5 a_{10} + a_6 a_9 - a_7 a_8 & -2(a_8 a_{11} - a_9 a_{10}) & -a_8 a_{15} + a_9 a_{14} + a_{10} a_{13} - a_{11} a_{12} \\ -a_0 a_7 + a_1 a_6 + a_2 a_5 - a_3 a_4 & -2(a_4 a_7 - a_5 a_6) & -a_4 a_{11} + a_5 a_{10} + a_6 a_9 - a_7 a_8 & -a_4 a_{15} + a_5 a_{14} + a_6 a_{13} - a_7 a_{12} \\ 2(a_0 a_3 - a_1 a_2) & a_0 a_7 - a_1 a_6 - a_2 a_5 + a_3 a_4 & a_0 a_{11} - a_1 a_{10} - a_2 a_9 + a_3 a_8 & a_0 a_{15} - a_1 a_{14} - a_2 a_{13} + a_3 a_{12} \end{pmatrix}$$

(23)

It is easy to show that character polynomial of $F_{12(34)}(\psi)$ is given by

$$\lambda^4 + a_3 \lambda^3 + a_2 \lambda^2 + a_1 \lambda + a_0$$

where $\quad a_3 = -H\lambda, \qquad\qquad\qquad a_2 = \frac{1}{4}H^2 + 2(L + 2M),$

$\qquad a_1 = -4D_{xt} - 6HM - HL, \qquad a_0 = L^2$. (24)

Therefore, the character polynomial of $F_{12(34)}(\psi)$ determine four independent algebraic SLOCC invariants $(H, L, M, D_{xt})$.

## 5. Conclusion

In summary, using revise definition of the square matrices $F(\psi)$, we can know if two *n*-qubit states are SLOCC equivalent then their square matrices $F(\psi)$ given above have the same character polynomial. Further, we find that just use one square matrices $F_{12(34)}(\psi)$, can we determine if two *n*-qubit states are SLOCC equivalent or not for four-qubit system. We expect this method will be useful in the determine SLOCC invariants for n-qubit states.

**Acknowledgments**

This work was supported by the foundation of Shannxi provincial Educational Department under Contract No. 15JK1668 and National Science Foundation of Shannxi Province Grant Nos 2017JQ6024.